\documentclass[pdflatex,sn-basic]{sn-jnl}

\jyear{2021}%

\theoremstyle{thmstyleone}%
\newtheorem{theorem}{Theorem}
\theoremstyle{thmstyletwo}%
\newtheorem{remark}{Remark}%

\theoremstyle{thmstylethree}%

\begin{document}

\title[Long-Term Returns Estimation of Leveraged Indexes and ETFs]{Long-Term Returns Estimation of Leveraged Indexes and ETFs}


\author{Hayden Brown\footnote{Department of Mathematics and Statistics, University of Nevada, Reno\\Email address: haydenb@nevada.unr.edu\\ORCID: 0000-0002-2975-2711}}




\abstract{Daily leveraged exchange traded funds amplify gains and losses of their underlying benchmark indexes on a daily basis. The result of going long in a daily leveraged ETF for more than one day is less clear. Here, bounds are given for the log-returns of a leveraged ETF when going long for more than just one day. The bounds are quadratic in the daily log-returns of the underlying benchmark index, and they are used to find sufficient conditions for outperformance and underperformance of a leveraged ETF in relation to its underlying benchmark index. Results show that if the underlying benchmark index drops 10+\% over the course of 63 consecutive trading days, and the standard deviation of the benchmark index's daily log-returns is no more than .015, then going long in a -3x leveraged ETF during that period gives a log-return of at least 1.5 times the log-return of a short position in the underlying benchmark index. Results also show promise for a 2x daily leveraged S\&P 500 ETF. If the average annual log-return of the S\&P 500 index continues to be at least .0658, as it has been in the past, and the standard deviation of daily S\&P 500 log-returns is under .0125, then a 2x daily leveraged S\&P 500 ETF will perform at least as well as the S\&P 500 index in the long-run.}

\keywords{Leveraged ETFs; Leveraged exchange traded funds; Inverse leveraged ETFs, Returns estimation}





\maketitle

\section{Introduction}\label{intro}
A daily leveraged exchange traded fund amplifies the daily return of its underlying benchmark index between adjusted closing prices (adjusted closing prices account for stock splits and dividends). For example, consider an underlying benchmark index returning 1\% between two consecutive adjusted closing prices. Then a daily leveraged ETF with leverage multiple 2x or -3x would return roughly 2\% or -3\%, respectively, depending on the expense ratio, fund management and whether it is trading at a premium or discount on the close. The results presented here aim at adressing the viability of going long in a daily leveraged ETF for more than one day. Their advantage over existing results in the literature is that they bound the leng-term log-return of a daily leveraged ETF, and they depend only on the daily leveraged ETF's expense ratio and the bounds, mean and standard deviation of the underlying benchmark index's daily log-returns.

It is now commonplace for investors to buy-and-hold an ETF tracking a large market index like the S\&P 500. Over a long period of time, like 40 years, investors are confident that such an ETF will provide a positive log-return that is favorable over most other available ETFs. But can the same be said about a daily leveraged S\&P 500 ETF? One of the main goals here is to determine when a daily leveraged ETF will outperform its underlying benchmark index in the long-run. 

When a stock or ETF looks overvalued, investors may take on a short position, hoping the price will drop in the near future. Alternatively, investors could buy-and-hold an inverse daily leveraged ETF during that period. Here, conditions are given indicating when the latter option offers a superior return.

There are toy examples where a portfolio that invests $L\%$ in a particular stock or ETF and $(1-L)\%$ in cash outperforms a portfolio that goes all in the stock or ETF. In general, this outperformance occurs when volatility is high enough. The results presented here show when this outperformance is impossible, with the goal being to validate the long-term portfolio that invests 100\% in a S\&P 500 ETF over a portfolio that reduces exposure to the S\&P 500. 

\subsection{Literature review}
Several empirical studies have measured the returns of daily leveraged ETFs over longer time spans than just one day. The general consensus is that leveraged ETFs track the leveraged multiple of their benchmark indexes' returns well in the short term but deviate in the long term. Since the deviations can be markedly negative, there is considerable risk associated with a long position in a leveraged ETF. 

Over time spans of 1, 3, 5 and 10 years, a 2x daily leveraged S\&P 500 ETF offers a moderate increase to expected return at the cost of a significant increase in standard deviation \cite{trainor2008leveraged}. Over time spans no longer than one month, 2x and -2x daily leveraged ETFs generally provide 2x and -2x, respectively, the return of the underlying benchmark index \cite{lu2009long}. For time spans longer than one month, serious deviations start to happen. Those deviations are attributed, in part, to the quadratic variation of the underlying benchmark index. On the other hand, \cite{bansal2015tracking} show that, for investment horizons of 1 calendar year from 1964 to 2013, the average difference between a leveraged S\&P 500 ETF's return and the underlying benchmark index's return multiplied by the leverage amount is greater than 0. So there is clearly potential for a daily leveraged S\&P 500 ETF to provide significant amplification of return over a calendar year. The results presented here complement these empirical findings by estimating returns of a daily leveraged ETF for investment horizons having any number of days.

Most theoretical results address a long-term position in a continuously leveraged ETF. Provided the underlying benchmark index follows a geometric Brownian motion, leveraged ETFs appear to cause value destruction in the long-run \cite{cheng2009dynamics}. At a minimum, leveraged ETFs do not achieve their leverage multiple in the long-run \cite{jarrow2010understanding}. The risk of a leveraged ETF is measured in \cite{leung2012leveraged}, and admissible leverage multiples are given accordingly. Using continuous leverage, \cite{giese2010risk} shows that dynamic adjustment of the leverage multiple based on market conditions leads to outperformance of the underlying benchmark index in the long-run. In reality, daily leveraged ETFs do not implement continuous leverage, so the forementioned results cannot be applied without assuming some level of error. The results presented here do not use continuous leverage to avoid this error. 

Other theoretical results address a long-term position in an ETF that is leveraged discretely in time. An approximation to the long-term return of a daily leveraged ETF is given by \cite{avellaneda2010path} for investment horizons of less than one year. It is based on the leverage multiple and the mean and variance of the underlying index's daily returns. Empirically, this approximation has been shown to be very accurate for quarterly horizons. However, it is not an upper or lower bound. The approximations presented here are advantageous because they are upper and lower bounds, which facilitates the provision of sufficient conditions for outperformance and underperformance of a daily leveraged ETF relative to its underlying benchmark index.

During the financial crisis from 2008 to 2009, daily leveraged ETFs did not generally meet their target multiple of daily returns, even on a daily basis \cite{shum2013leveraged}. Similar finding are in \cite{tang2013solving}. These errors can be attributed to management and trading premiums/discounts, and the effect is a reduction in the magnification of daily returns. For example, 2x and -2x daily leveraged S\&P 500 ETFs were more like 1.9x and -1.9x daily leveraged ETFs during the financial crisis. These errors are not considered here because their randomness is difficult to incorporate into theoretical results. However, the results can account for such errors, to some extent, with an increased expense ratio. 

Based on simulation of 3x and -3x daily leveraged S\&P 500 ETFs, it appears that a combination of volatility and market condition (sideways, upward-trending or downward-trending) of the S\&P 500 index determines long-term performance of the leveraged ETF \cite{charupat2022understanding}. The results presented here provide a theoretical foundation for these simulation-based findings.

Daily leveraged ETFs are certainly popular, but it appears that their present use by institutions is leading to poor performance relative to portfolios that avoid daily leveraged ETFs \cite{devault2021blessing}. In other words, recent attempts by institutions to time the market with their leverage ETF holdings are backfiring. The results presented here are aimed at providing further guidance on when a leveraged ETF is worth having in a portfolio.

\subsection{Main results}\label{imain}
Lower and upper bounds are given for the log-return of a daily leveraged index over $n$ consecutive trading days. The bounds are expressed quadratically in terms of the daily log-returns of the underlying benchmark index. In particular, the bounds are of the form $n(am_2+bm_1+c)$, where $m_2$ is the average squared daily log-return of the benchmark index, $m_1$ is the average daily log-return of the benchmark index, and $a$, $b$ and $c$ are constants. The results cover a range of leverage multiples, including the popular -3x, -2x, 2x and 3x. 

\subsection{Applications}
Sufficient conditions are given for the log-return of a daily leveraged index or ETF to be some multiple, $L_0$, of the log-return of its underlying benchmark index, over $n$ consecutive trading days. Here, ETFs are distinguished from indexes because they have expense ratios. To simplify notation, let $R^L_{n,r}$ denote the log-return of a daily leveraged ETF after $n$ consecutive trading days, with leverage multiple $L$ and expense ratio $r$. Now the goal of applications can be expressed more concisely: to provide sufficient conditions for $R^L_{n,r}$ to be at least or at most $L_0R^1_{n,0}$. 

First, thresholds are given for $m_1/m_2$, indicating when $R^L_{n,r}$ is at least or at most $L_0R^1_{n,0}$. Here, $m_1$ and $m_2$ are as in Section \ref{imain}. Special attention is given to the thresholds for $1<L$ because 2x and 3x leverage multiples are so popular.

Let $s$ denote the standard deviation of the underlying benchmark index's daily log-returns. For $L>1$ and $L_0<L$, an upper bound is given on $s$, indicating when $R^L_{n,r}\geq L_0R^1_{n,0}$. Taking $L=2,3$ and $L_0=0,1$ is especially important for practical reasons, because the upper bound on $s$ indicates when a 2x or 3x daily leveraged ETF will have a non-negative log-return or perform at least as well as its underlying benchmark index. If the average annual log-return of the S\&P 500 index continues to be at least .0658, as it has been in the past, daily percentage changes between adjusted closing prices are at least -20\%, and the standard deviation of daily log-returns is under .0125, then a 2x daily leveraged ETF will perform at least as well as the S\&P 500 index in the long-run.

For $L<L_0<0$, an upper bound is given on $s$, indicating when $R^L_{n,r}\geq L_0R^1_{n,0}$. The focus here is on $L_0=-1$ because then the latter inequality indicates when a daily inverse leveraged ETF performs at least as well as a short position in its underlying benchmark index. For example, results show that if the benchmark index drops 10+\% over the course of 63 consecutive trading days, daily percentage changes between adjusted closing prices are at most 15\%, and $s\leq.015$, then going long in a -3x leveraged ETF during that period gives a log-return of at least 1.5 times the log-return of a short position in the benchmark index. Furthermore, if the 10+\% drop happens faster, then the 1.5 multiple of log-returns can be achieved with even larger $s$.

For $0<L<1$, an upper bound is given on $s$, indicating when $R^L_{n,0}\leq R^1_{n,0}$. Note that a log-return of $R^L_{n,0}$ can be achieved via daily rebalancing with $L\%$ in the benchmark index and $(1-L)\%$ in cash. Interestingly, this theory easily extends from daily leverage to longer periods like weekly leverage, where rebalancing occurs weekly, and quarterly leverage, where rebalancing occurs quarterly. If the S\&P 500 continues to have an average annual log-return of at least .0658, then the standard deviation of its daily, weekly, monthly, quarterly, semi-annual or annual log-returns would have to exceed .02, .04, .08, .15, .2 or .35, respectively, for a portfolio rebalancing daily, weekly, monthly, quarterly, semi-annually or annually, respectively, with $.64<L<1$, to outperform the benchmark $L=1$ in the long-run. It seems unlikely for this level of volatility to persist in the long-run, so maintaining a $L:(1-L)$ portfolio in the benchmark and cash with $64<L<1$ is not advised under any standard rebalancing schedule.

\subsection{Organization}
Section \ref{prelim} lays out the notation and framework for the returns of daily leveraged indexes and ETFs. Section \ref{mainresults} provides main results, and Section \ref{applications} applies those results. Data used in applications is described in Section \ref{data}. Section \ref{conclusion} provides closing remarks, including a discussion of related future research ideas. Last, \ref{secA1} provides proofs of the theorems stated in Section \ref{mainresults}.

\section{Preliminaries}\label{prelim}
Let $C_i$ denote the adjusted closing price of trading day $i$ for a particular stock market index I. Then $\{C_i\}_{i=0}^n$ is a sequence of adjusted closing prices for $n+1$ consecutive trading days. Note that adjusted closing prices account for dividends and stock splits, but not inflation. For example, suppose a stock has a closing price of \$100 on day 1, a closing price of \$98 on day 2, and a dividend distribution of \$1 on day 2. Then the adjusted closing prices will be \$100 on day 1 and \$99 on day 2. Suppose there is a 2-for-1 stock split on day 3 and the closing price on day 3 is \$49. Then the adjusted closing price of day 3 will be $\$99=2\cdot\$49+\$1$. 

Let $X_i=C_{i}/C_{i-1}-1$ for $i=1,...,n$. Then $\{100\cdot X_i\}_{i=1}^n$ is the sequence of $n$ percentage changes between adjusted closing prices. Observe that 
\begin{equation*}
\prod_{i=1}^n(1+X_i)=\frac{C_n}{C_0}.
\end{equation*}

Denote the daily leveraged version of I as LxI, where L indicates the amount of leverage. For example, 3xI indicates the index tracking I with 3x daily leverage. The closing prices of LxI are given by 
\begin{equation*}
C_i^L:=C_0\cdot\prod_{k=1}^i(1+LX_k), \quad i=0,...,n.
\end{equation*}
So the log-returns realized by going long in LxI from the close of trading day $0$ to the close of trading day $n$ are given by
\begin{equation*}
\log \frac{C_n^L}{C_0}=\sum_{i=1}^n\log(1+LX_i).
\end{equation*}
Note that here, $\log$ refers to the natural logarithm. Let $Y_i=\log(1+X_i)$ for $i=1,...,n$. Then $Y_i$ is the log-return for day $i$, and 
\begin{equation*}
\log\frac{C_n}{C_0}=\sum_{i=1}^nY_i,\quad \log \frac{C_n^L}{C_0}=\sum_{i=1}^n\log(1+L(\exp Y_i-1)).
\end{equation*}
To shorten notation, let
\begin{equation*}
m_1=\frac{1}{n}\sum_{i=1}^nY_i,\quad m_2=\frac{1}{n}\sum_{i=1}^nY_i^2,\quad s=\sqrt{\frac{\sum_{i=1}^n(Y_i-m_1)^2}{n}}.
\end{equation*}

Denote the ETF version of LxI as LxI$_r$, where $r$ is the annual expense ratio, compounded on a daily basis. Assuming 252 trading days in a year, the log-return of LxI$_r$ after $n$ days is given by
\begin{equation*}
R_{n,r}^L:=\log\frac{C_n^L}{C_0}-n\log\Big(1+\frac{r}{252}\Big).
\end{equation*}

\section{Main Results}\label{mainresults}
Theorems \ref{T1}, \ref{T2a}, \ref{T2b} and \ref{T3} provide lower and upper bounds for the log-return of LxI. The bounds are expressed quadratically in terms of the $Y_i$. Theorem \ref{T1} covers $L>1$, Theorems \ref{T2a} and \ref{T2b} cover $0<L<1$, and Theorem \ref{T3} covers $L<0$.
\begin{theorem}
Fix $L>1$ and $\log(1-L^{-1})<y_0<y_1$. Then, provided $y_0\leq Y_i\leq y_1$ for $i=1,...,n$, log-returns of LxI from the close of trading day $0$ to the close of trading day $n$ are bounded as follows:
\begin{equation*}
\sup_{y_0<y}n(a_0m_2+b_0m_1+c_0)\leq\log \frac{C_n^L}{C_0}\leq\inf_{\log(1-L^{-1})<y<y_1}n(a_1m_2+b_1m_1+c_1),
\end{equation*}
where
\begin{equation*}
\begin{split}
&a_k=\frac{1}{y-y_k}\Bigg(\frac{\log\frac{1+L(\exp y_k-1)}{1+L(\exp y-1)}}{y-y_k}+\frac{L\exp y}{1+L(\exp y-1)}\Bigg),\\
&b_k=\frac{L\exp y}{1+L(\exp y-1)}-2a_ky,\quad c_k=\log(1+L(\exp y_k-1))-a_ky_k^2-b_ky_k.
\end{split}
\end{equation*}
\label{T1} 
\end{theorem}

\begin{remark}
In Theorem \ref{T1}, the requirements $\log(1-L^{-1})<y_0$ and $Y_i\geq y_0$ for $i=1,...,n$ guarantee that $X_i>L^{-1}$ for each $i$. This, in turn, makes each daily leveraged return $1+LX_i$ well-defined (i.e. non-negative).
\end{remark}

\begin{remark}
In Theorem \ref{T1}, having $y=0$ and $y_0<0<y_1$ simplifies the expressions for $a_k,\ b_k,$ and $c_k$ considerably. In particular, 
\begin{equation*}
a_k=\frac{1}{y_k}\Bigg(\frac{\log(1+L(\exp y_k-1))}{y_k}-L\Bigg),\quad b_k=L,\quad c_k=0.
\end{equation*}
\label{c0}
\end{remark}

\begin{theorem}
Fix $0<L<1$ and $y_0<y_1<\log(L^{-1}-1)$. Then, provided $y_0\leq Y_i\leq y_1$ for $i=1,...,n$, log-returns of LxI from the close of trading day $0$ to the close of trading day $n$ are bounded as follows:
\begin{equation*}
\sup_{y_0<y<\log(L^{-1}-1)}n(a_0m_2+b_0m_1+c_0)\leq\log \frac{C_n^L}{C_0}\leq\inf_{y<y_1}n(a_1m_2+b_1m_1+c_1),
\end{equation*}
where $a_k,\ b_k$ and $c_k$ are as in Theorem \ref{T1}.
\label{T2a} 
\end{theorem}

\begin{theorem}
Fix $0<L<1$ and $\log(L^{-1}-1)<y_0<y_1$. Then, provided $y_0\leq Y_i\leq y_1$ for $i=1,...,n$, log-returns of LxI from the close of trading day $0$ to the close of trading day $n$ are bounded as follows:
\begin{equation*}
\sup_{\log(L^{-1}-1)<y<y_1}n(a_1m_2+b_1m_1+c_1)\leq\log \frac{C_n^L}{C_0}\leq\inf_{y_0<y}n(a_0m_2+b_0m_1+c_0),
\end{equation*}
where $a_k,\ b_k$ and $c_k$ are as in Theorem \ref{T1}.
\label{T2b} 
\end{theorem}

\begin{theorem}
Fix $L<0$ and $y_0<y_1<\log(1-L^{-1})$. Then, provided $y_0\leq Y_i\leq y_1$ for $i=1,...,n$, log-returns of LxI from the close of trading day $0$ to the close of trading day $n$ are bounded as follows:
\begin{equation*}
\sup_{y<y_1}n(a_1m_2+b_1m_1+c_1)\leq\log \frac{C_n^L}{C_0}\leq\inf_{y_0<y<\log(1-L^{-1})}n(a_0m_2+b_0m_1+c_0),
\end{equation*}
where $a_k,\ b_k$ and $c_k$ are as in Theorem \ref{T1}.
\label{T3} 
\end{theorem}

\begin{remark}
In Theorem \ref{T3}, the requirements $y_1<\log(1-L^{-1})$ and $Y_i\leq y_1$ for $i=1,...,n$ guarantee that $X_i<-L^{-1}$ for each $i$. This, in turn, makes each daily leveraged return $1+LX_i$ well-defined (i.e. non-negative).
\end{remark}

\begin{remark}
For any $L\in\mathbb{R}\setminus[0,1]$, there is also the linear upper bound 
\begin{equation*}
\log \frac{C_n^L}{C_0}\leq Lnm_1=L\log \frac{C_n}{C_0},
\end{equation*}
which holds provided each $Y_i$ satisfies 
\begin{equation*}
\frac{L}{|L|}Y_i>\frac{L}{|L|}\log(1-L^{-1}).
\end{equation*}
Thus, the log-return of LxI can never exceed $L$ times the log-return of I. Note that this upper bound follows the fact that $\log(1+Lx)\leq L\log(1+x)$ where the logarithms are well-defined. 

For $L\in[0,1]$, the upper bound just described is a lower bound, i.e.
\begin{equation*}
L\log \frac{C_n}{C_0}=Lnm_1\leq\log \frac{C_n^L}{C_0}.
\end{equation*}
\end{remark}

\section{Applications}\label{applications}
Applications first use Theorem \ref{T1} and Remark \ref{c0} to provide thresholds for $m_1/m_2$ indicating when LxI has a log-return that is at least or at most $L_0$ times the log-return of I. The focus is on $1<L$ and $L_0<L$, but similar thresholds exist for arbitrary $L$ and $L_0$ based on Theorems \ref{T2a}, \ref{T2b} and \ref{T3}. Note that $m_1/m_2$ is akin to a Sharpe ratio, since $m_1$ and $m_2$ denote means of the $Y_i$ and $Y_i^2$, respectively. 

Next, main results are used to to provide sufficient conditions for the log-return of LxI$_r$ to be at least $L_0$ times the log-return of I. For practical reasons, the focus is on two cases: ($1<L$, $L_0<L$) and ($L<L_0<0$). The former case is aimed at indicating when a leveraged ETF like 2xI$_r$ or 3xI$_r$ outperforms I. The latter case is aimed at indicating when an inverse leveraged ETF like -2xI$_r$ or -3xI$_r$ outperforms a short position in I.

Last, Theorems \ref{T2a} and \ref{T2b} are used to provide sufficient conditions for the log-return of LxI to be at most the log-return of I. Here, the focus is on $0<L<1$ because then the log-return of LxI can be achieved by maintaining a portfolio with $L\%$ in I and $(1-L)\%$ in cash. Interestingly, this theory easily extends from daily leverage to longer periods like weekly leverage, where rebalancing occurs weekly, and quarterly leverage, where rebalancing occurs quarterly.

\subsection{Data}\label{data}
Applications use the average annual real log-return of the S\&P Composite Index from 1871 to 2020, which is .0658. Here, S\&P Composite Index refers to three indexes: Cowles and Associates from 1871 to 1926, Standard \& Poor 90 from 1926 to 1957 and Standard \& Poor 500 from 1957 to 2020. The Cowles and Associates and S\&P 90 indexes are backward extensions of the S\&P 500 index used to extrapolate a longer term average annual real log-return of the S\&P 500 index. The data was taken from \url{http://www.econ.yale.edu/~shiller/data.html} and is collected for easy access at \url{https://github.com/HaydenBrown/Investing}. For an overview of the S\&P 500, see \url{https://www.spglobal.com/spdji/en/indices/equity/sp-500/}. Relevant variables from the data are described below. 
\begin{table}[h]
\begin{center}
\caption{Data variable descriptions}
\begin{tabular}{ |c|l| } 
\hline
\textbf{Notation} & \textbf{Description} \\
 \hline
 $P$ & average monthly close of the S\&P composite index \\ 
 \hline
 $D$ & dividend per share of the S\&P composite index \\ 
 \hline
 $J$ & January consumer price index \\ 
 \hline
\end{tabular}
\end{center}
\end{table}
Inflation and dividend adjusted (i.e. real) annual returns are computed using the consumer price index, the S\&P Composite Index price and the S\&P Composite Index dividend. Use the subscript $k$ to denote the $k$th year of $J$, $P$ and $D$. Then the real return for year $k$ is given by $((P_{k+1}+D_k)/P_k)\cdot(J_k/J_{k+1})$. Note that the average annual total log-return (adjustment for dividends but not inflation) of the S\&P Composite Index from 1871 to 1926 is greater than .0658, since inflation has been far more common than deflation in the past. 

\subsection{Thresholds for $\frac{m_1}{m_2}$}\label{thresh}
\begin{figure} 
  \includegraphics[width=\linewidth]{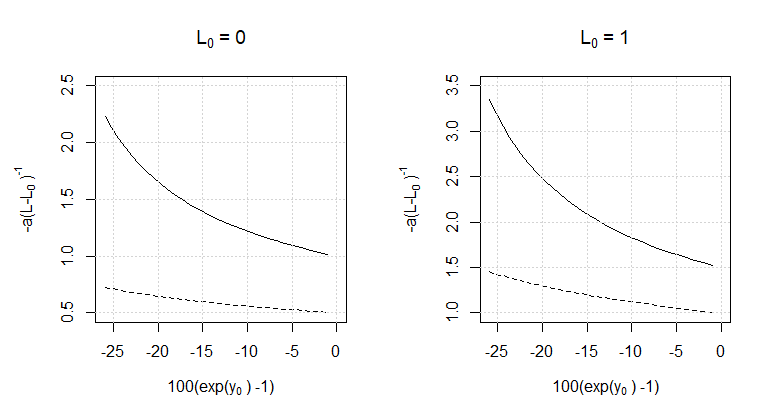}
  \caption{Illustrates $\frac{-a_0}{L-L_0}$ for various $y_0,\ L$ and $L_0$. The dashed line represents $L=2$, and the solid line represents $L=3$.}
  \label{fig:threshold}
\end{figure}

Let $1<L$, $L_0<L$ and $\log(1-L^{-1})<y_0<y_1$. Observe that $L_0\log(C_n/C_0)=L_0nm_1$. It follows from Theorem \ref{T1} and Remark \ref{c0} that $L_0\log(C_n/C_0)\leq\log(C_n^L/C_0)$, provided $L_0m_1\leq a_0m_2+Lm_1$ and $y_0\leq Y_i\leq y_1$ for $i=1,...,n$. If at least one $Y_i$ is non-zero, then $m_2$ is positive and 
\begin{equation}
L_0m_1\leq a_0m_2+Lm_1\iff \frac{-a_0}{L-L_0}\leq \frac{m_1}{m_2}.
\label{RSR}
\end{equation}
Of special interest are the cases $L_0=0$ and $L_0=1$. When $L_0=0$, satisfaction of \eqref{RSR} indicates LxI has a non-negative log-return. When $L_0=1$, satisfaction of \eqref{RSR} indicates LxI has a log-return that is at least the log-return of I. Moreover, it is not hard to see that when $L_0=1$, satisfaction of \eqref{RSR} indicates the log-return of LxI$_r$ is at least the log-return of 1xI$_r$. Figure \ref{fig:threshold} illustrates the threshold $-a_0(L-L_0)^{-1}$ for various $y_0,\ L$ and $L_0$. The horizontal axis is in terms of $100(\exp(y_0)-1)$ for the sake of interpretation. Observe that $Y_i\geq y_0$ if and only if $100X_i\geq100(\exp(y_0)-1)$. So if $Y_i\geq y_0$ for $i=1,...,n$, then $100(\exp(y_0)-1)$ indicates the lower bound on the daily percentage changes between adjusted closing prices. 

Using similar logic, $L_0\log(C_n/C_0)\geq\log(C_n^L/C_0)$, provided $y_0\leq Y_i\leq y_1$ for $i=1,...,n$ and
\begin{equation*}
\frac{-a_1}{L-L_0}\geq \frac{m_1}{m_2}.
\end{equation*}
Using Theorems \ref{T2a}, \ref{T2b} and \ref{T3}, similar thresholds for $m_1/m_2$ can be made for arbitrary $L$ and $L_0$, indicating when $L_0\log(C_n/C_0)$ is at least or at most $\log(C_n^L/C_0)$.

\subsection{Outperformance of Daily Leveraged Indexes and ETFs}\label{outp}
In general, Theorems \ref{T1}, \ref{T2a}, \ref{T2b} and \ref{T3} can be used to provide sufficient conditions for the log-return of LxI$_r$ to be at least or at most some multiple, denoted $L_0$, of the log-returns of I, i.e.
\begin{equation*}
L_0\log(C_n/C_0)\leq R_{n,r}^L\quad\text{or}\quad L_0\log(C_n/C_0)\geq R_{n,r}^L. 
\end{equation*}
Rather than detail said sufficient conditions for the many cases that arise when considering general $L$ and $L_0$, it is more worthwhile to focus on a few cases of practical interest. 

The goal here is to provide sufficient conditions for the log-return of LxI$_r$ to be at least $L_0$ times the log-return of I, i.e.
\begin{equation}
L_0\log(C_n/C_0)\leq R_{n,r}^L.
\label{goal1}
\end{equation}
Furthermore, only two cases of $L$ and $L_0$ are considered for practical interest:
\begin{enumerate}[(i)]
\item $1<L$ and $L_0<L$,
\item $L<L_0<0$.
\end{enumerate}

\subsubsection{Case (i)}
Fix $\log(1-L^{-1})<y_0<y_1$, and suppose $y_0\leq Y_i\leq y_1$ for $i=1,...,n$. Recall that $L_0\log(C_n/C_0)=L_0nm_1$. By Theorem \ref{T1}, \eqref{goal1} follows, provided 
\begin{equation}
L_0m_1\leq\sup_{y_0<y}a_0m_2+b_0m_1+c_0-\log\Big(1+\frac{r}{252}\Big).
\label{sc1}
\end{equation}
Some algebra (and the fact that $a_0<0$) reveals that \eqref{sc1} is equivalent to 
\begin{equation*}
s\leq\sup_{y_0<y}\sqrt{-m_1^2+\frac{L_0-b_0}{a_0}\cdot m_1-\frac{c_0-\log(1+r/252)}{a_0}}.
\end{equation*}
Recall that $s$ denotes the standard deviation of the $Y_i$. 

\begin{figure} 
  \includegraphics[width=\linewidth]{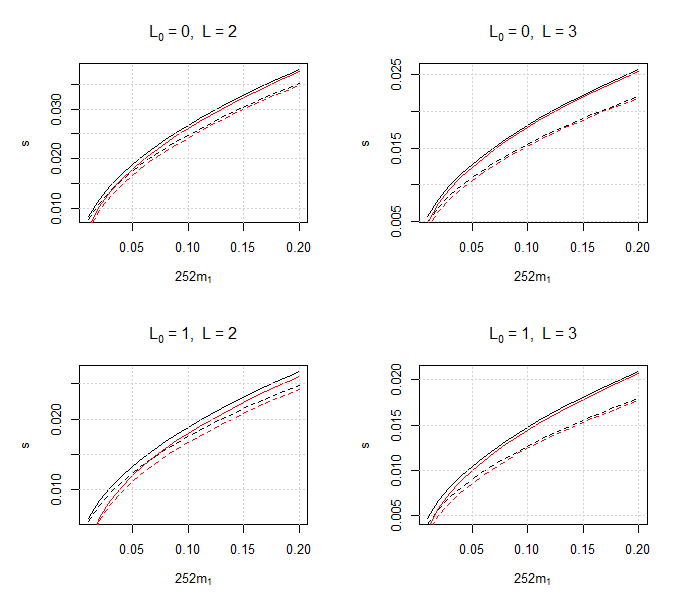}
  \caption{Illustrates \eqref{seq} for various $y_0,\ L,\ L_0$ and $r$. The dashed line represents $y_0=\log(1-.2)$ (minimum daily percentage change is -20\%), and the solid line represents $y_0=\log(1-.1)$ (minimum daily percentage change is -10\%). Black indicates $r=0$ and red indicates $r=.0095$ (expense ratio is .95\%).}
  \label{fig:Lpos1}
\end{figure}

\begin{figure} 
  \includegraphics[width=\linewidth]{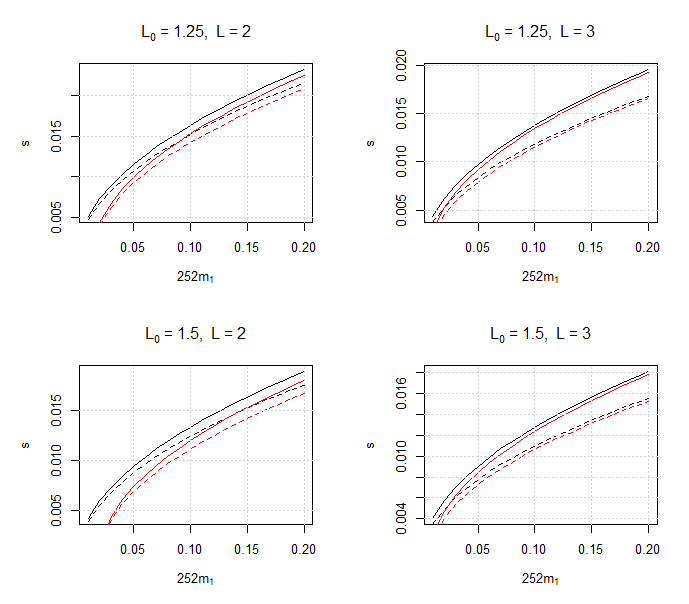}
  \caption{Illustrates \eqref{seq} for various $y_0,\ L,\ L_0$ and $r$. The dashed line represents $y_0=\log(1-.2)$ (minimum daily percentage change is -20\%), and the solid line represents $y_0=\log(1-.1)$ (minimum daily percentage change is -10\%). Black indicates $r=0$ and red indicates $r=.0095$ (expense ratio is .95\%).}
  \label{fig:Lpos2}
\end{figure}

For various average annual log-return of I (i.e. $252m_1$), Figures \ref{fig:Lpos1} and \ref{fig:Lpos2} show the standard deviation, $s$, of daily log-returns that satisfies 
\begin{equation}
s=\sup_{y_0<y}\sqrt{-m_1^2+\frac{L_0-b_0}{a_0}\cdot m_1-\frac{c_0-\log(1+r/252)}{a_0}}.
\label{seq}
\end{equation}
If the standard deviation of daily log-returns is less than or equal to what is shown in Figures \ref{fig:Lpos1} and \ref{fig:Lpos2}, then \eqref{sc1} is satisfied, which, in turn, implies \eqref{goal1}. Observe how the impact of the expense ratio on \eqref{seq} is exaggerated as $L_0$ increases. Note that Figures \ref{fig:Lpos1} and \ref{fig:Lpos2} approximate the supremum over $y>y_0$ with a fine mesh. Interestingly, the $y$ that produces the approximate supremum is close to 0 in each case. Recall that the lower bound of Theorem \ref{T1} is constructed to be especially close to the actual leveraged log-return when the $Y_i$ are close to $y$. Having $y\approx0$ makes the lower bound especially accurate when the $Y_i$ are close to 0. If a quadratic lower bound is selected from Theorem \ref{T1}, but the $y$ is selected based on intuition rather than taking the supremum, then it makes sense to choose $y\approx0$, because daily log-returns should have a mean close to 0 in the long-run. 

\begin{figure} 
  \includegraphics[width=\linewidth]{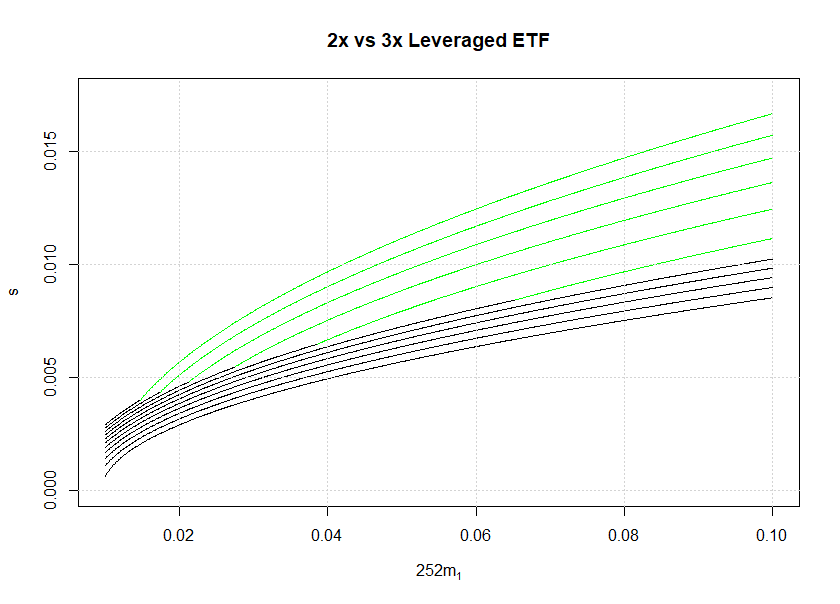}
  \caption{Illustrates \eqref{seq} for $y_0=\log(1-.2)$, $L=2,3$, $L_0=1,1.1,...,1.9,2$ and $r=.0095$. The top curve is for $L_0=1$, the curve below that is for $L_0=1.1$, ..., and the bottom curve is for $L_0=2$ (there are 11 curves total, some colored green and black, some just black). Green indicates $L=2$ produced the larger value for \eqref{seq} than $L=3$, and that value is plotted in green. Black indicates $L=3$ produced the larger value for \eqref{seq} than $L=2$, and that value is plotted in black.}
  \label{fig:2v3}
\end{figure}

Figure \ref{fig:2v3} shows that, for $y_0=\log(1-.2)$, $L_0\geq1.6$ and $r=.0095$ , $L=3$ has a larger upper bound for $s$. If one is trying to achieve at least 1.6 times the log-return of I with a leveraged ETF, it appears that 3xI$_r$ can do so while allowing a higher standard deviation of the daily log-returns of I, when compared to 2xI$_r$. Note, the word ``appears" is used because the main results only provide sufficient conditions for outperformance. For $L_0<1.6$, the situation can be reversed, depending on the value of $252m_1$. For example, if one is trying to achieve 1.4 times the log-return of I, and $252m_1>.04$, then it appears that 2xI$_r$ can do so while allowing a higher standard deviation of the daily log-returns of I. Furthermore, if one is trying to achieve at least the log-return of I, then it appears that 2xI$_r$ is far more forgiving than 3xI$_r$, in terms of the allowed standard deviation of the daily log-returns of I.

Based on Section \ref{data}, the average annual real log-return of the S\&P Composite Index from 1871 to 2020 is .0658 (recall that real log-return indicates adjustment for dividends and inflation). Assuming this trend continues, inflation occurs in the long-run, daily percentage changes between adjusted closing prices are at least -20\%, and the standard deviation of daily log-returns is under .0125, Figure \ref{fig:Lpos1} indicates that a 2x daily leveraged S\&P 500 ETF with a .95\% expense ratio will perform at least as good as the benchmark S\&P 500 index in the long-run. Moreover, the situation only improves when there is persistent inflation, in which case the upper bound on the standard deviation of daily log-returns increases. On the other hand, if one takes the position that the benchmark S\&P 500 index is unbeatable in the long-run, then it is necessary for the standard deviation of daily log-returns to be at least .0125, provided an average annual real log-return of 0.0658, persistent inflation, and daily percentage changes between adjusted closing prices of at least -20\%. This means that if the benchmark S\&P 500 index is unbeatable in the long-run, average annual log-returns continue to be at least 0.0658, and daily percentage changes between adjusted closing prices are not too extreme, then daily log-returns must be quite volatile.

\subsubsection{Case (ii)}
Fix $y_0<y_1<\log(1-L^{-1})$, and suppose $y_0\leq Y_i\leq y_1$ for $i=1,...,n$. Recall that $L_0\log(C_n/C_0)=L_0nm_1$. By Theorem \ref{T3}, \eqref{goal1} follows, provided 
\begin{equation}
L_0m_1\leq\sup_{y<y_1}a_1m_2+b_1m_1+c_1-\log\Big(1+\frac{r}{252}\Big).
\label{sc2}
\end{equation}
Some algebra (and the fact that $a_1<0$) reveals that \eqref{sc2} is equivalent to 
\begin{equation*}
s\leq\sup_{y<y_1}\sqrt{-m_1^2+\frac{L_0-b_1}{a_1}\cdot m_1-\frac{c_1-\log(1+r/252)}{a_1}}.
\end{equation*}

\begin{figure} 
  \includegraphics[width=\linewidth]{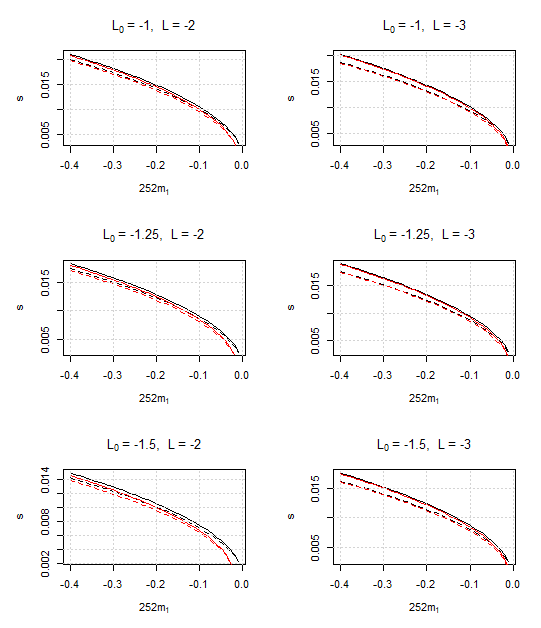}
  \caption{Illustrates \eqref{seq} for various $y_1,\ L,\ L_0$ and $r$. The dashed line represents $y_1=\log(1+.15)$ (maximum daily percentage change is 15\%), and the solid line represents $y_1=\log(1+.1)$ (maximum daily percentage change is 10\%). Black indicates $r=0$ and red indicates $r=.0095$ (expense ratio is .95\%).}
  \label{fig:Lneg}
\end{figure}

\begin{figure} 
  \includegraphics[width=\linewidth]{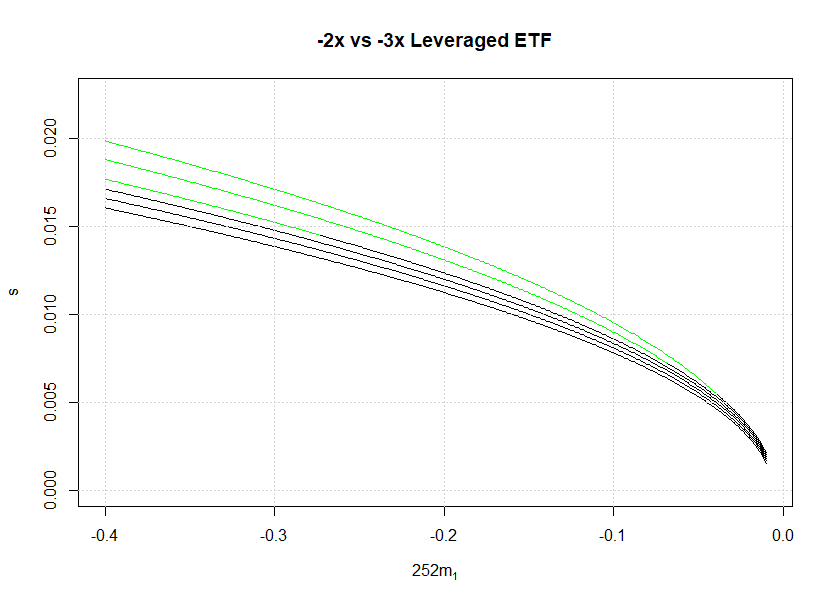}
  \caption{Illustrates \eqref{seq} for $y_0=\log(1+.15)$, $L=2,3$, $L_0=-1,-1.1,...,-1.5$ and $r=.0095$. The top curve is for $L_0=-1$, the curve below that is for $L_0=-1.1$, ..., and the bottom curve is for $L_0=-1.5$ (there are 6 curves total, some colored green and black, some just black). Green indicates $L=-2$ produced the larger value for \eqref{seq} than $L=-3$, and that value is plotted in green. Black indicates $L=-3$ produced the larger value for \eqref{seq} than $L=-2$, and that value is plotted in black.}
  \label{fig:n2vn3}
\end{figure}

For various average annual log-returns of I (i.e. $252m_1$), Figure \ref{fig:Lneg} shows the standard deviation, $s$, of daily log-returns that satisfies 
\begin{equation}
s=\sup_{y<y_1}\sqrt{-m_1^2+\frac{L_0-b_1}{a_1}\cdot m_1-\frac{c_1-\log(1+r/252)}{a_1}}.
\label{seq2}
\end{equation}
If the standard deviation of daily log-returns is less than or equal to what is shown in Figure \ref{fig:Lneg}, then \eqref{sc2} is satisfied, which, in turn, implies \eqref{goal1}. Unlike case (i), Figure \ref{fig:Lneg} shows that \eqref{seq2} is approximately the same for a variety of $L_0$ and $L$. If one is confident of a downturn and daily log-returns of I are not too volatile, then -2xI$_r$ and -3xI$_r$ are both good options for magnifying returns. Looking at Figure \ref{fig:n2vn3}, it appears that -3xI$_r$ is superior (in terms of allowable standard deviation of the $Y_i$) when looking to magnify log-returns of $I$ by $L_0\geq-1.3$. For $L_0<-1.3$, it appears that -2xI$_r$ is superior in the same sense, provided $252m_1$ is sufficiently negative. 

\subsection{Underperformance of Leveraged Indexes}
The goal here is to provide sufficient conditions for the log-return of LxI to be at most the log-return of I, i.e.
\begin{equation}
\log(C_n/C_0)\geq \log(C_n^L/C_0).
\label{goal2}
\end{equation}
For practical interest, only $0<L<1$ are considered. Note that leveraged ETFs are not considered here because the return of LxI can be achieved with ease by a private investor via rebalancing. Of course, there are fees and taxes associated with the rebalancing that lead to decreased returns. Thus, if \eqref{goal2} holds, then the investor's log-return will be less than or equal to the log-return of I.

First generalize the framework outlined in Section \ref{prelim}. Instead of $\{C_i\}_{i=0}^n$ being a sequence of adjusted closing prices for $n+1$ consecutive trading days, take it to be a subsequence coming from $\{\mathcal{C}_i\}_{i=0}^N$ $(n\leq N)$, a sequence of adjusted closing prices for $N+1$ consecutive trading days. 

Next, require rebalancing at the close of each day associated with $C_i$ such that $(1-L)\%$ is in cash and $L\%$ is in I. The resulting log-return between trading days associated with $C_{i-1}$ and $C_{i}$ is then 
\begin{equation*}
\log[(1-L)+L(C_i/C_{i-1})]=\log[(1-L)+L(X_i+1)]=\log(1+LX_i)=Y_i.
\end{equation*}
So this schedule of rebalancing achieves the same leveraged returns used in the main results, and Theorems \ref{T2a} and \ref{T2b} can be applied.

Fix $y_0<y_1<\log(L^{-1}-1)$, and suppose $y_0\leq Y_i\leq y_1$ for $i=1,...,n$. Recall that $\log(C_n/C_0)=nm_1$. By Theorem \ref{T2a}, \eqref{goal2} follows, provided 
\begin{equation}
\inf_{y<y_1}a_1m_2+b_1m_1+c_1\leq m_1.
\label{sc2a}
\end{equation}
Some algebra (and the fact that $a_1>0$) reveals that \eqref{sc2a} is equivalent to 
\begin{equation*}
s\leq\inf_{y<y_1}\sqrt{-m_1^2+\frac{1-b_1}{a_1}\cdot m_1-\frac{c_1}{a_1}}.
\end{equation*}

\begin{figure} 
  \includegraphics[width=\linewidth]{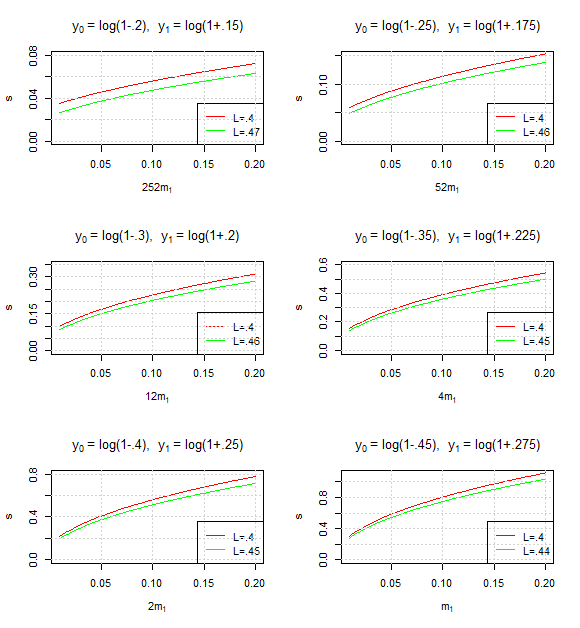}
  \caption{Illustrates \eqref{seq2a}. Horizontal axes measure the average annual log-return of I. The coefficient on $m_1$ indicates the average number of times rebalancing occurs each year. In particular, $252m_1$ is for daily rebalancing, $52m_1$ is for weekly rebalancing, $12m_1$ is for monthly rebalancing, $4m_1$ is for quarterly rebalancing, $2m_1$ is for semi-annual rebalancing, and $m_1$ is for annual rebalancing. $y_0$ and $y_1$ are adjusted for each rebalancing schedule. In each plot, the green curve indicates the $L$ satisfying $y_1=\log(L^{-1}-1)$, rounded down to the nearest hundredth. The red curve is there to show which direction \eqref{seq2a} goes in relation to the green curve as $L$ is decreased. Note that smaller values for $L$, like $.01$, produce a \eqref{seq2a} that is significantly higher than the green curve, so much so that it would not show up on the given plots.}
  \label{fig:Lunderhalf}
\end{figure}

For various rebalancing schedules and average annual log-returns of I, Figure \ref{fig:Lunderhalf} shows the standard deviation, $s$, of daily log-returns that satisfies 
\begin{equation}
s=\inf_{y<y_1}\sqrt{-m_1^2+\frac{1-b_1}{a_1}\cdot m_1-\frac{c_1}{a_1}}.
\label{seq2a}
\end{equation}
If the standard deviation of daily log-returns is less than or equal to what is shown in Figure \ref{fig:Lunderhalf}, then \eqref{sc2a} is satisfied, which, in turn, implies \eqref{goal2}. 

Fix $\log(L^{-1}-1)<y_0<y_1$, and suppose $y_0\leq Y_i\leq y_1$ for $i=1,...,n$. Recall that $\log(C_n/C_0)=nm_1$. By Theorem \ref{T2b}, \eqref{goal2} follows, provided 
\begin{equation}
\inf_{y_0<y}a_0m_2+b_0m_1+c_0\leq m_1.
\label{sc2b}
\end{equation}
Some algebra (and the fact that $a_0>0$) reveals that \eqref{sc2b} is equivalent to 
\begin{equation*}
s\leq\inf_{y_0<y}\sqrt{-m_1^2+\frac{1-b_0}{a_0}\cdot m_1-\frac{c_0}{a_0}}.
\end{equation*}

\begin{figure} 
  \includegraphics[width=\linewidth]{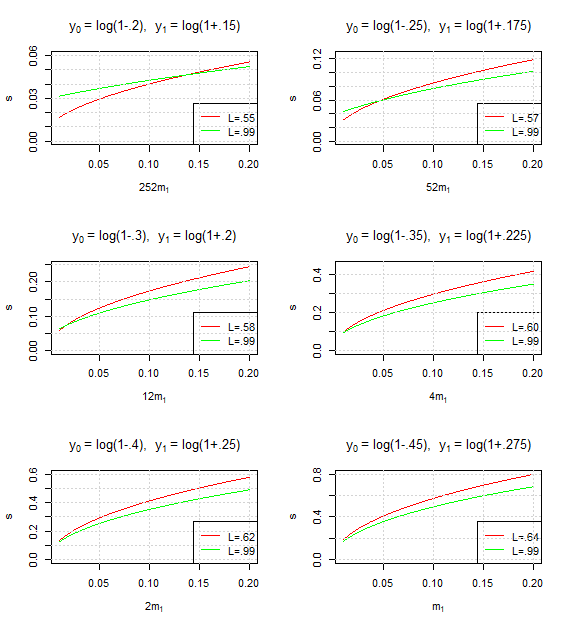}
  \caption{Illustrates \eqref{seq2b}. Horizontal axes measure the average annual log-return of I. The coefficient on $m_1$ indicates the average number of times rebalancing occurs each year. In particular, $252m_1$ is for daily rebalancing, $52m_1$ is for weekly rebalancing, $12m_1$ is for monthly rebalancing, $4m_1$ is for quarterly rebalancing, $2m_1$ is for semi-annual rebalancing, and $m_1$ is for annual rebalancing. $y_0$ and $y_1$ are adjusted for each rebalancing schedule. In each plot, the green curve indicates the $L$ satisfying $\log(L^{-1}-1)=y_0$, rounded up to the nearest hundredth. The red curve is there to show which direction \eqref{seq2b} goes in relation to the green curve as $L$ is increased.}
  \label{fig:Lunder1}
\end{figure}

For various rebalancing schedules and average annual log-returns of I, Figure \ref{fig:Lunder1} shows the standard deviation, $s$, of daily log-returns that satisfies 
\begin{equation}
s=\inf_{y_0<y}\sqrt{-m_1^2+\frac{1-b_0}{a_0}\cdot m_1-\frac{c_0}{a_0}}.
\label{seq2b}
\end{equation}
If the standard deviation of daily log-returns is less than or equal to what is shown in Figure \ref{fig:Lunder1}, then \eqref{sc2b} is satisfied, which, in turn, implies \eqref{goal2}. 

\begin{figure} 
  \includegraphics[width=\linewidth]{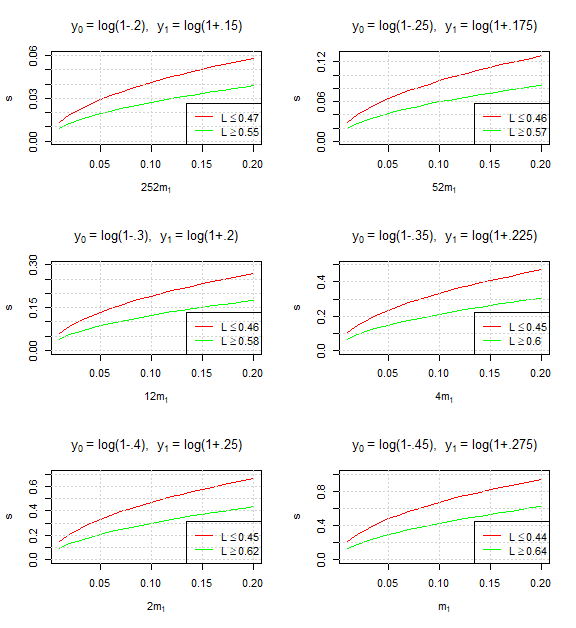}
\caption{The red curve shows the minimum of \eqref{seq2a} over $L\in S_1$, and the green curve shows the minimum of \eqref{seq2b} over $L\in S_2$. Horizontal axes measure average the annual log-return of I. The coefficient on $m_1$ indicates the average number of times rebalancing occurs each year. In particular, $252m_1$ is for daily rebalancing, $52m_1$ is for weekly rebalancing, $12m_1$ is for monthly rebalancing, $4m_1$ is for quarterly rebalancing, $2m_1$ is for semi-annual rebalancing, and $m_1$ is for annual rebalancing. $y_0$ and $y_1$ are adjusted for each rebalancing schedule.}
  \label{fig:Lbetter}
\end{figure}

Fix $y_0<y_1$ and suppose $y_0\leq Y_i\leq y_1$ for $i=1,...,n$. Figure \ref{fig:Lbetter} provides upper bounds on $s$ that imply \eqref{goal2} for a range of $L$. Let
\begin{equation*}
S_1=\{L\in S:\ y_1<\log(L^{-1}-1)\},\quad S_2=\{L\in S:\ \log(L^{-1}-1)<y_0\},
\end{equation*}
where $S=\{.01,.02,...,.99\}$. If $L\in S_1$ and $s$ is less than or equal to the its value on the red curve, then \eqref{goal2} holds. If $L\in S_2$ and $s$ is less than or equal to the its value on the green curve, then \eqref{goal2} holds. Figure \ref{fig:Lbetter} indicates that \eqref{goal2} holds even when the daily, weekly, monthly, quarterly, semi-annual or annual log-returns of I are quite volatile. Unfortunately, Theorems \ref{T2a} and \ref{T2b} do not cover all $L$ between 0 and 1. No upper bound on $s$ is given for $L\in[(1+\exp y_1)^{-1},\ (1+\exp y_0)^{-1}]$. However, as shown in Figure \ref{fig:Lbetter}, $S_1$ and $S_2$ contain most of the $L$ in $S$. $S_1$ contains $L$ less than .5, and $S_2$ contains $L$ greater than .5, with both missing some $L$ close to .5. For practical purposes, an investor is likely choosing between going all in on I or maintaining a portfolio of $L\%$ in I and $(1-L)\%$ in cash, with $L$ close to 1 (i.e. $L\in S_2$). The goal here would be to dial down exposure to I, and slightly increase return as a result. However, Figure \ref{fig:Lbetter} shows that there would need to be substantial volatility in the daily log-returns of I for this slight increase in return to be possible. For example, take I to be the S\&P 500. If the S\&P 500 continues to have an average annual log-return of at least .0658, then the standard deviation of its daily, weekly, monthly, quarterly, semi-annual or annual log-returns would have to exceed .02, .04, .08, .15, .2 or .35, respectively, for a portfolio that dials down exposure in I to be viable. It seems unlikely for this level of volatility to persist in the long-run, so maintaining a $L:(1-L)$ portfolio in I and cash with $L\in S_2$ is not advised.

\section{Conclusions \& Further Research}\label{conclusion}
The quadratic bounds considered here are useful because they provide simple thresholds indicating when a leveraged index or ETF will underperform or outperform its underlying benchmark index. Using similar methods, cubic (or even quartic) bounds can be constructed. A cubic bound would be tighter than a quadratic bound because of the additional degree of freedom. However, a cubic bound uses the average $Y_i^3$, which could be difficult to interpret. 

The methods used here cover most leverage multiples of practical interest. However, no quadratic bounds are given for $L\in[(1+\exp y_1)^{-1},\ (1+\exp y_0)^{-1}]$, where $y_0$ and $y_1$ are the lower and upper bounds on the $Y_i$. Issues arise because now the derivative of $\log(1+L(\exp x-1))$ has a change in concavity between $y_0$ and $y_1$. This change in concavity makes it so the methods used to prove Theorems \ref{T1}, \ref{T2a}, \ref{T2b} and \ref{T3} cannot be directly applied to $L\in[(1+\exp y_1)^{-1},\ (1+\exp y_0)^{-1}]$. Perhaps some modification of the methods used here can be used to deal with $L\in[(1+\exp y_1)^{-1},\ (1+\exp y_0)^{-1}]$.

\begin{appendix}
\section{Proofs}\label{secA1}
\section*{Proof of Theorem \ref{T1}}
First the lower bound is shown to hold. Define $p,f:[y_0,\infty)\to\mathbb{R}$ such that $p(x)=a_0x^2+b_0x+c_0$ and $f(x)=\log(1+L(\exp x-1))$. Let $y>y_0$. The values for $a_0,\ b_0$ and $c_0$ given in the statement of the theorem are found by solving the system $f(y_0)=p(y_0)$, $f(y)=p(y)$ and $\frac{df}{dx}(y)=\frac{dp}{dx}(y)$. From here, the goal is to show $p(x)\leq f(x)$ for all $x\in[y_0,\infty)$. Then the result will follow because 
\begin{equation*}
n(am_2+bm_1+c)=\sum_{i=1}^np(Y_i),\quad \log \frac{C_n^L}{C_0}=\sum_{i=1}^nf(Y_i).
\end{equation*}

By Rolle's Theorem, there is a $x^*\in(y_0,y)$ such that $\frac{df}{dx}(x^*)-\frac{dp}{dx}(x^*)=0$. Next observe that 
\begin{equation*}
\frac{d^3}{dx^3}[f(x)-p(x)]=\frac{(1-L)L(1-L(\exp x+1))\exp x}{(1+L(\exp x-1))^3}.
\end{equation*}
It follows that $\frac{d}{dx}[f(x)-p(x)]$ is strictly convex, since $L>1$ and $x\geq y_0>\log(1-L^{-1})$. By strict convexity, the only zeros of $\frac{d}{dx}[f(x)-p(x)]$ are $x^*$ and $y$. Moreover, $x^*$ is a local max, and $y$ is a local min. Combining this with the fact that $y_0$ and $y$ are zeros of $f-p$ reveals that $f-p\geq0$.

For the upper bound, redefine $p,f:(\log(1-L^{-1}),y_1]\to\mathbb{R}$ such that $p(x)=a_1x^2+b_1x+c_1$ and $f(x)=\log(1+L(\exp x-1))$. Let $y\in(\log(1-L^{-1}),y_1)$. The values for $a_1,\ b_1$ and $c_1$ given in the statement of the theorem are found by solving the system $f(y_1)=p(y_1)$, $f(y)=p(y)$ and $\frac{df}{dx}(y)=\frac{dp}{dx}(y)$. From here, the goal is to show $p(x)\geq f(x)$ for all $x\in(\log(1-L^{-1}),y_1]$. The rest of the proof follows similar logic as was used to verify the lower bound.
\begin{flushright}$\square$\end{flushright}

\section*{Proof of Theorems \ref{T2a} and \ref{T2b}}
The proof is very similar to that of Theorem \ref{T1}. Just observe that now the function $f(x)=\log(1+L(\exp x-1))$ is well-defined on $\mathbb{R}$, and it has $\frac{df}{dx}$ strictly convex on $(-\infty,\log(L^{-1}-1))$ and strictly concave on $(\log(L^{-1}-1),\infty)$.
\begin{flushright}$\square$\end{flushright}

\section*{Proof of Theorem \ref{T3}}
The proof is very similar to that of Theorem \ref{T1}. Just observe that now the function $f(x)=\log(1+L(\exp x-1))$ is well-defined on $(-\infty,\log(1-L^{-1}))$, and it has $\frac{df}{dx}$ strictly concave.
\begin{flushright}$\square$\end{flushright}

\end{appendix}


\bibliography{sn-bibliography}

\end{document}